\documentclass[twocolumn,astrosymb]{aastex631}

\usepackage{multirow}
\usepackage{placeins}
\usepackage{gensymb}
\usepackage{float}
\usepackage{color}
\usepackage{bm}
\usepackage{soul}
\usepackage{hyperref}

\shorttitle{Satellites of Milky Way Analogs in R{\sc omulus}25}
\shortauthors{Van Nest et al.}

\graphicspath{{./}{figures/}}

\received{\today}
\revised{tomorrow}
\accepted{day after tomorrow}
\submitjournal{ApJ}

\begin{document}

\title{The Role of Mass and Environment on Satellite distributions around Milky Way analogs in the {\sc Romulus25} simulation}

\correspondingauthor{Jordan Van Nest}
\email{jdvannest@ou.edu}

\author{Jordan Van Nest}
\affiliation{Homer L. Dodge Department of Physics \& Astronomy, University of Oklahoma, 440 W. Brooks St., Norman, OK 73019, USA}

\author{Ferah Munshi}
\affiliation{Department of Physics \& Astronomy, George Mason University, 4400 University Drive, MSN: 3F3 Fairfax, VA 22030-4444, USA}

\author{Charlotte Christensen}
\affiliation{Department of Physics, Grinnell College, 1116 8th Ave., Grinnell, IA 50112, USA}

\author{Alyson M. Brooks}
\affiliation{Department of Physics \& Astronomy, Rutgers, The State University of New Jersey, 136 Frelinghuysen Road, Piscataway, NJ 08854, USA}
\affiliation{Center for Computational Astrophysics, Flatiron Institute, 162 Fifth Ave, New York, NY 10010, USA}

\author{Michael Tremmel}
\affiliation{Astronomy Department, Yale University, P.O Box 208120, New Haven, CT 06520, USA}
\affiliation{Physics Department, University College Cork, T12 K8AF Cork, Ireland}

\author{Thomas R. Quinn}
\affiliation{Astronomy Department, University of Washington, Box 351580, Seattle, WA 98195-1580, USA}

\begin{abstract}
We study satellite counts and quenched fractions for satellites of Milky Way analogs in {\sc Romulus25}, a large-volume cosmological hydrodynamic simulation. Depending on the definition of a Milky Way analog, we have between 66 and 97 Milky Way analogs in {\sc Romulus25}, a 25 Mpc per-side uniform volume simulation. We use these analogs to quantify the effect of environment and host properties on satellite populations. We find that the number of satellites hosted by a Milky Way analog increases predominantly with host stellar mass, while environment, as measured by the distance to a Milky Way-mass or larger halo, may have a notable impact in high isolation.  Similarly, we find that the satellite quenched fraction for our analogs also increases with host stellar mass, and potentially in higher-density environments. These results are robust for analogs within 3 Mpc of another Milky Way-mass or larger halo, the environmental parameter space where the bulk of our sample resides. We place these results in the context of observations through comparisons to the Exploration of Local VolumE Satellites and Satellites Around Galactic Analogs surveys. Our results are robust to changes in Milky Way analog selection criteria, including those that mimic observations. Finally, as our samples naturally include Milky Way-Andromeda pairs, we examine quenched fractions in pairs vs isolated systems.  We find potential evidence, though not conclusive, that pairs, defined as being within 1 Mpc of another Milky Way-mass or larger halo, may have higher satellite quenched fractions. 
\end{abstract}


\section{Introduction}
\label{sec:intro}

The satellites of the Milky Way and its neighbors in the Local Group, thanks to their proximity, have often served as our basis of understanding satellite and dwarf galaxy formation and evolution. In the past decades there has been an explosion in our understanding of satellites around our own Milky Way \citep[][and references within]{Mateo98,Koposov08,Simon19,DrlicaWagner20} and Andromeda \citep[][and references within]{Ibata14,Martin16,McConnachie18}. Further, in the age of ultrafaint galaxy detection, the low surface brightness end of the Milky Way's satellite distribution continues to grow \citep[e.g.][]{DrlicaWagner15,Koposov15,Simon19}. As we continue to discover fainter objects nearby, the question of the Milky Way's uniqueness becomes an important one. Applying what we learn locally to the Universe at large would not be appropriate if the Local Group could be considered ``atypical''. 

To test for any potential discrepancy, surveys such as the ``Satellites Around Galactic Analogs'' \citep[SAGA;][]{SAGAI,SAGAII} and ``Exploration of Local VolumE Satellites'' \cite[ELVES;][]{Carlsten20,Carlsten21a,Carlsten21b,Carlsten22} study the satellite distributions of galaxies similar to our own, placing the Milky Way in a broader, cosmological context. The SAGA survey is an ongoing effort to compile spectroscopically complete satellite luminosity functions of 100 Milky Way analogs with distances between 20 and 40 Mpc, providing vastly improved statistics for the bright end of these satellite distributions (down to $M_R$=-12.3).  In complement to the SAGA survey's probing of distant Milky Way-like systems, the ELVES survey seeks to fully map the satellite distributions of the hosts within the Local Volume fully ($<12$ Mpc) down to $M_V$=-9.

Working in tandem, SAGA and ELVES will provide a better understanding of both what a ``typical'' Milky Way-like halo will look like and what influences an environment like the Local Volume can impart. The SAGA survey has found that the luminosity function of the Milky Way is consistent with their observations of other systems, but that the host-to-host scatter in the number of satellites is large \citep{SAGAII}. SAGA also finds that the total number of satellites in a system correlates with the host's $K$-band luminosity. Similar to SAGA, the ELVES survey finds that satellite abundance correlates with host mass and that the Milky Way is typical for its mass.  However, \citet{Carlsten21b} find that the observed luminosity functions of local hosts are typically ``flatter'' than predicted by the cosmological model; the stellar-to-halo mass relation tends to underpredict bright satellites and overpredict faint ones, a result found also by \citet{SAGAI}. These results highlight the power of a larger sample of galaxies and their satellites to provide context for understanding satellite dwarf galaxies.

One of the most interesting discrepancies to be highlighted so far is that the quenched fraction of Local Group (Milky Way and Andromeda) satellites is not in agreement with SAGA's results: the SAGA sample exhibits lower quenched fractions than those found in the Local Group.  On the other hand, the ELVES survey finds higher quenched fractions amongst the Local Volume than in the SAGA sample, though still not as high as the Local Group.  Although \citet{SAGAII} carefully attempt to quantify incompleteness in the SAGA survey, it remains an open question whether SAGA may be missing faint, red, or low surface brightness satellites which would be predominantly quenched \citep{Carlsten22, Font22}, or whether the Local Group is a true outlier in terms of quenched satellite fraction.

In general, various simulations of Milky Way-like galaxies tend to find good agreement in their resulting quenched satellite fractions, lying somewhere between the Local Group and Local Volume fractions \citep{Akins21, Engler21, Karunakaran21, Samuel22}.  These simulations generally find that galaxies that infall into a host Milky Way with stellar masses above $M_* \sim 10^8$ $M_\odot$ are better able to retain their gas and continue star forming for extended periods.  On the other hand, galaxies with stellar masses below $10^8$ $M_\odot$ instead tend to experience ram pressure stripping that strips gas and quenches their star formation (SF), and the quenching time scales can often be quite short \citep[$<2$ Gyr;][]{Wetzel15,Simpson18,Simons20,Akins21}. These results lead to high predicted quenched satellite fractions as luminosity decreases. 

On the theoretical front, many analyses use zoom-in simulations of a handful of Milky Way analogs \citep[e.g.,][]{Akins21, Samuel22}, though \citet{Font22} use the {\sc artemis} suite of 24 cosmological Milky Way-mass zooms to interpret the ELVES and SAGA results.  \citet{Font22} find that applying a surface brightness limit to the {\sc artemis} satellites can bring the quenched fractions and radial distributions into line with the SAGA results, suggesting that SAGA is missing faint surface brightness galaxies.  Fainter surface brightnesses correlate with more quenching at a fixed luminosity in {\sc artemis}, and thus bias the SAGA results if true.  On the other hand, \citet{Engler23} found that a surface brightness cut could not bring the TNG50 satellite quenched fractions fully into agreement with SAGA, though it did bring the simulation and observational results more into line.  \citet{Engler23} were able to use TNG50, a 50 Mpc on-a-side uniform cosmological volume, to study a larger sample of Milky Way analogs and look for statistical trends.  In this work, we use {\sc Romulus25}, a 25 Mpc on-a-side uniform cosmological volume with comparable resolution to TNG50, to study similar trends.  We particularly focus on the questions of how host mass and large-scale environment impact both satellite counts and quenched fractions for our simulated Milky Way analogs. 

The paper is outlined as follows.  We begin in Section \ref{sec:sim} by describing the R{\sc omulus}25 simulation. In Section \ref{sec:sample} we outline our various methods for identifying Milky Way analogs, as well as their satellites. In Section \ref{sec:results} we present our primary results, focusing on the general size of the satellite populations and their quenched fractions. We then discuss and summarize our results in Sections \ref{sec:discussion} \& \ref{sec:conclusions}, respectively.

\section{Simulation}
\label{sec:sim}

For this work, we use the {\sc Romulus25} simulation \citep{Tremmel17}. {\sc Romulus25} was run with {\sc ChaNGa} \citep{Changa}, which includes standard physics modules previously used in {\sc GASOLINE} \citep{wadsley04,wadsley08,wadsley17} such as a cosmic UV background \citep{Haardt12} including self-shielding \citep{pontzen08}, SF, `blastwave' supernova (SN) feedback \citep{Stinson06}, and low-temperature metal cooling \citep{Bromm01}. {\sc ChaNGa} implements an updated smoothed particle hydrodynamics (SPH) routine that uses a geometric mean density in the SPH force expression, allowing for the accurate simulation of shearing flows with Kelvin-Helmholtz instabilities \citep{wadsley17}. Finally, a time-dependent artificial viscosity and an on-the-fly time-step adjustment \citep{saitoh09} system allow for more realistic treatment of weak and strong shocks \citep{wadsley17}.

R{\sc omulus}25 assumes a $\Lambda$ cold dark matter ($\Lambda$CDM) model with cosmological parameter values following results from Planck \citep[$\Omega_0=0.3086$, $\Lambda=0.6914$, $h=0.6777$, and $\sigma_8=0.8288$;][]{planck16}. The simulation has a Plummer equivalent force softening of 250 pc (a spline softening of 350 pc is used, which converges to a Newtonian force at 700 pc). Unlike many similar cosmological runs, the dark matter particles were {\it oversampled} relative to gas particles, such that the simulation was run with initially $3.375$ times more dark matter particles than gas. This increased dark matter resolution allows for the ability to track the dynamics of supermassive black holes within galaxies \citep{tremmel15}. The result is a dark matter particle mass of $3.39 \times 10^5$$M_{\odot}$ and gas particle mass of $2.12 \times 10^5$$M_{\odot}$. These relatively low dark matter particle masses decrease numerical effects resulting from two-body relaxation and energy equipartition, which occur when particles have significantly different masses, both of which can affect the structure of simulated galaxies \citep[e.g.,][]{ludlow19}.  {\sc Romulus25} has been shown to reproduce important galaxy and supermassive black hole scaling relations \citep{Tremmel17, ricarte19, Sharma22, Sharma23}.

\subsection{SF and gas cooling}\label{sec:sf}

Gas cooling at low temperatures is regulated by metal abundance as in \citet{eris11}, as well as SPHs that include both thermal and metal diffusion as described in \citet{shen10} and \citet{governato15} (thermal and metal diffusion coefficients set to 0.3, see \citet{Tremmel17,Tremmel19} for an in-depth discussion). SF and associated feedback from SNe are crucial processes that require subgrid models in cosmological simulations like {\sc Romulus25}.  Following \citet{Stinson06}, SF is regulated with parameters that encode SF efficiency in dense gas, couple SN energy to the interstellar medium (ISM), and specify the physical conditions required for SF. These parameters were calibrated using several dozen zoom-in simulations of dwarf to Milky Way-mass galaxies \citep{Tremmel17} and are as follows.

\begin{enumerate}
\setlength\itemsep{1em}
\item The normalization of the SF efficiency, $c_{\rm SF} = 0.15$, and formation timescale, $\Delta t = 10^6$ yr, are both used to calculate the probability $p$ of creating a star particle from a gas particle that has a dynamical time $t_{\text{dyn}}$

\begin{equation}
p =\frac{m_{\rm gas}}{m_{\rm star}}(1 - e^{-c_{\rm SF} \Delta t /t_{\text{dyn}}}).
\end{equation}
 
\item The fraction of SN energy coupled to the ISM, $\epsilon_{\rm SN} = 0.75$.

\item Minimum density, $n_{\star} = 0.2$ cm$^{-3}$, and maximum temperature, $T_\star = 10^4$ K, thresholds beyond which cold gas is allowed to form stars.
 
 \end{enumerate}

Star particles form with a mass of $6\times10^4$ $M_{\odot}$, or 30\% the initial gas particle mass. R{\sc omulus}25 assumes a Kroupa initial mass function \citep{Kroupa01} with associated metal yields and SN rates. Feedback from SNe uses the `blastwave' implementation \citep{Stinson06}, with thermal energy injection and a cooling shutoff period approximating the `blastwave' phase of SN ejecta when cooling is inefficient.

\subsection{Halo Identification}

Amiga Halo Finder \citep[AHF;][]{AHF} was applied to R{\sc omulus}25 to identify dark matter halos, subhalos, and the baryonic content within. AHF uses a spherical top-hat collapse technique \citep{BryanNorman98} to calculate each halo's virial radius ($R_{vir}$) and mass ($M_{vir}$). Halos are considered resolved if their virial mass is at least $3\times10^9$ $M_\odot$ at $z=0$. This corresponds to a dark matter particle count of $\sim10^4$, and a stellar mass of at least $10^7$ $M_{\odot}$ (star particle count of $\sim150$). Following \citet{Munshi13}, stellar masses were scaled by a factor of 0.5 as a photometric correction for a better comparison to values inferred from typical observational techniques, and all magnitudes use the Vega zero-point.

\begin{deluxetable*}{c|c|c|c|c}
\tablenum{1}
\tablewidth{0pt}
\tablecaption{A Summary of Our Samples of Milky Way Analogs and Satellites}
\tablehead{
\colhead{(1) Analog Criteria} & \colhead{(2) Analog Radius} & \colhead{(3) $N_{MW}$} & \colhead{(4) $N_{Sats}$} & \colhead{(5) max($N_{Sat}$)}
}
\startdata
$M_{\text{vir}}$ & \multirow{3}{4em}{$R_{\text{vir}}$} & 67 & 138 & 8 \\
$M_{*}$ &  & 97 & 210 & 13 \\
$M_{K}$+Env.  &  & 77 & 148 & 13 \\
\hline
$M_{\text{vir}}$ & \multirow{3}{4em}{300 kpc} & 66 & 125 & 6 \\
$M_{*}$ &  & 90 & 171 & 7 \\
$M_{K}$+Env. (SAGA II) &  & 77 & 137 & 6 \\
\enddata
\tablecomments{(1) The criteria for identifying Milky Way analogs; (2) the virial radius of the Milky Way analog for the purpose of identifying satellites; (3) the total number of Milky Way analogs; (4) the total number of satellites with $M_*$ $>$ 10$^7$ $M_{\odot}$; and (5) the largest number of satellites hosted by a single Milky Way analog.}

\label{tab:sample}
\end{deluxetable*}

\begin{figure}
    \centering
    \includegraphics[width=\linewidth]{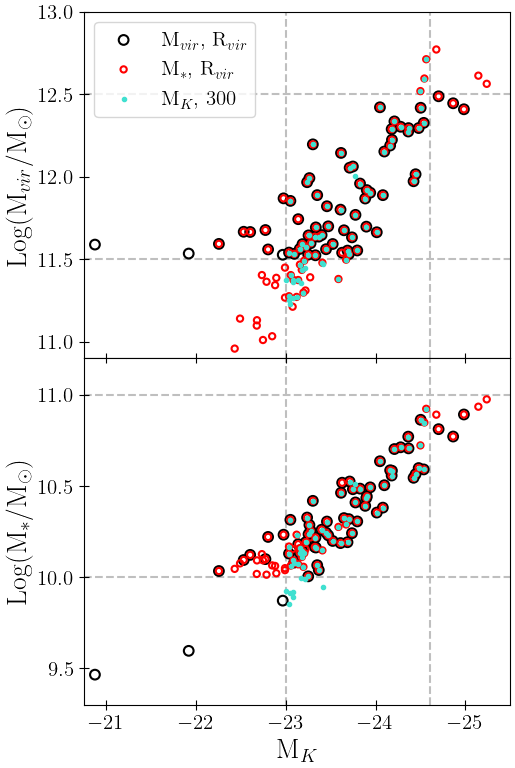}
    \caption{Virial and stellar masses plotted against $K$-band magnitudes for three of our Milky Way analog samples. The dotted lines denote the mass and magnitude cuts used in our samples. The samples diverge at the different boundaries, and even within the boundaries there are analogs that exist in some definitions but not others. }
    \label{fig:defcomp}
\end{figure}

\begin{figure*}
\gridline{
    \fig{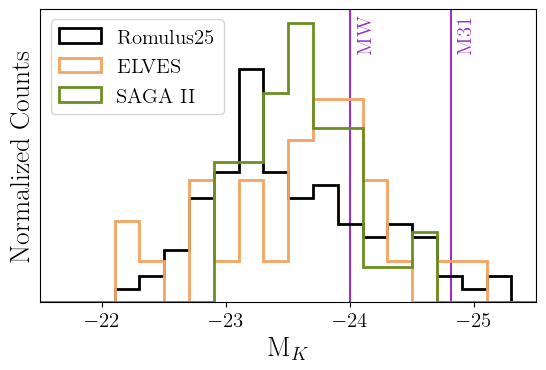}{.474\textwidth}{(a) Hosts} 
    \fig{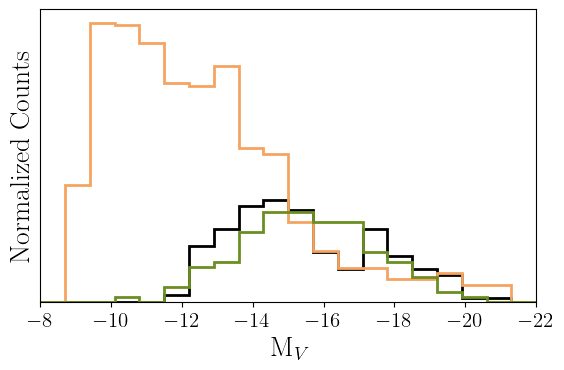}{.49\textwidth}{(b) Satellites}
}
\caption{Normalized histograms of (a) hosts in $K$-band magnitude and (b) satellites in $V$-band magnitude for our $M_*$ with simulated $R_{\text{vir}}$  sample. We make direct comparisons to SAGA II \citep{SAGAII} and ELVES \citep{Carlsten22} data, with the Milky Way and M31 values taken from the latter. The ELVES satellites are weighted according to their likelihood measurements. For a fair comparison, the satellite distributions in (b) are all normalized to the samples' number of satellites brighter than $M_V$=-14, the approximate completeness limit for R{\sc omulus}25.}
\label{fig:HostComp}
\end{figure*}

\section{Analog and Satellite Identification}
\label{sec:sample}

There is no concrete definition of what constitutes a Milky Way analog; observational surveys like SAGA and ELVES make sample cuts using $K$-band magnitudes as proxies for stellar mass, while simulations have access to more exact values for halo properties such as stellar mass and virial radius. In this work, we select samples of Milky Way analogs according to three different criteria sets in order to test if the selection criteria can influence the resultant satellite distribution. Our samples are defined as follows.
\begin{itemize}
    \item \textit{A general $M_{\text{vir}}$ restriction}: any halo where $10^{11.5} < $$M_{\text{vir}}$/$M_{\odot} < 10^{12.5}$.
    \item \textit{A general $M_*$ restriction}: any galaxy where $10^{10}<$$M_*$/$M_{\odot} < 10^{11}$. This corresponds to the host stellar mass range outlined in Section 2.1.2 of SAGA II \citep{SAGAII}. A stellar mass of 10$^{10}$ $M_\odot$ also corresponds to the lower limit of the ELVES survey \citep{Carlsten22}.
    \item \textit{An $M_{k}$ + Environmental restriction}: any galaxy where $-24.6<$$M_K<-23$. Additionally, no neighbor within 300 kpc can have $M_{K,\text{neighbor}} <$ $M_{K,\text{MW}}-1.6$. This corresponds to the $K$-band magnitude cut and environmental restrictions from SAGA II \citep{SAGAII}.
\end{itemize}

We also explore two different ways to identify a satellite galaxy. First, we consider galaxies within the host's virial radius down to a stellar mass of 10$^7$ $M_\odot$, the resolution limit for R{\sc omulus}25. This corresponds to an magnitude limit of $M_R\approx -12.6$. We note that the magnitude limit for the SAGA survey is $M_R=-12.3$ (though they have four satellites below this limit; see \citet{SAGAII}), so our samples do not probe the lowest-mass regions of the SAGA or ELVES sample spaces. In addition, we perform a selection where satellites are identified by being within 300 kpc of a Milky Way analog, rather than the analog's virial radius, as a more direct comparison to the SAGA and ELVES surveys. We note, however, that these surveys use 2D projected distances while in this work we use true 3D distances. In the event that a satellite is hosted by multiple analogs, it is ascribed to the most massive host. Any satellites that fall into the criteria of a Milky Way analog are not included in the satellite distribution. As a final step, any analogs that host a ``satellite'' more massive than themselves are removed from consideration. This cut is partially responsible for the slight variation in the number of Milky Way analogs under the same criteria when switching between the $R_{\text{vir}}$ and 300 kpc satellite identifications. Our sample of Milky Way analogs and satellites are summarized for each criteria set in Table \ref{tab:sample}.

Figure \ref{fig:defcomp} shows the three Milky Way analog samples that we focus on in this work: $M_{\text{vir}}$ and $M_*$ with $R_{\text{vir}}$ and $M_K$ with 300 kpc. While the samples largely overlap, we find that none of them are simple subsets of the others. As they approach the boundaries of the selection cuts, the samples diverge from one another. For example, the stellar mass sample probes virial masses below the virial mass cut, and vice versa. This is the result of natural scatter within the stellar-halo mass relation, which was shown in \citet{Tremmel17} to match observations \citep{Moster13,Kravtsov18}. Within the overlapping regions of the criteria, there are galaxies considered Milky Way analogs in some samples but not others. This occurs as a result of the environmental criteria in the SAGA sample, which could remove analogs that are still within the $K$-band magnitude limits.

In Figure \ref{fig:HostComp}, we compare the normalized distributions of hosts and satellites from our largest sample, $M_*$ with simulated $R_{\text{vir}}$ (in order to encapsulate the full magnitude range of our samples), to data from SAGA II and ELVES \citep{SAGAII,Carlsten22}. We note that the ELVES satellites are weighted according to their likelihood estimates ($P_{\text{sat}}$ in Table 9 of \cite{Carlsten22}), so each satellite adds its likelihood as a count rather than 1. In panel (a), we see that our hosts' span in $K$-magnitude space matches well with the ELVES sample, while the SAGA II sample (by definition) resides in $-24.6<M_K<-23$. The peaks of the host distributions are in good agreement as well, though we note our peak is at a slightly dimmer magnitude than the observational data. In panel (b), we see that our satellite distribution is in very good agreement with the SAGA II data, though we have an interesting lack of satellites at $M_K\approx-17$. The ELVES data probe much dimmer satellites (due to the difference in observational limits), but when only considering satellites brighter than $M_K=-12$, the ELVES sample is still more concentrated at low-mass satellites when compared to SAGA II and R{\sc omulus}25. This is consistent with ELVES finding steeper luminosity functions (fewer high-mass satellites and more low-mass ones) in their sample when compared to SAGA, and might also contribute to the different quenched fractions found by the two surveys (see Section \ref{sec:res_q} for discussion).

\begin{figure}
    \centering
    \includegraphics[width=\linewidth]{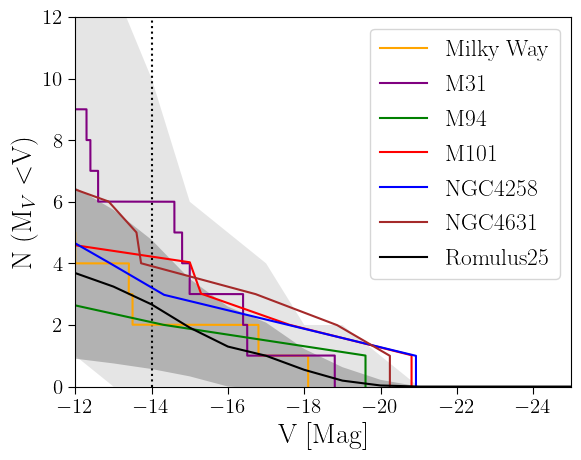}
    \caption{The $V$-band satellite luminosity function for our Milky Way analog sample under the $M_*$ with simulated $R_{\text{vir}}$ criteria. The black line and dark gray region represent the mean and single standard deviation of our sample, respectively, while the outer light gray region encompasses our entire sample. We compare to the Milky Way and M31 \citep{SAGAI}, M94 \citep{Smercina18}, M101 \citep{Bennet19}, and NGC 4258 and NGC 4631 \citep{Carlsten21b}. The dotted vertical line marks the approximate completeness limit for R{\sc omulus}25. Our sample is in good agreement with these observations.}
    \label{fig:lumfunc}
\end{figure}

\begin{figure*}
\gridline{
    \fig{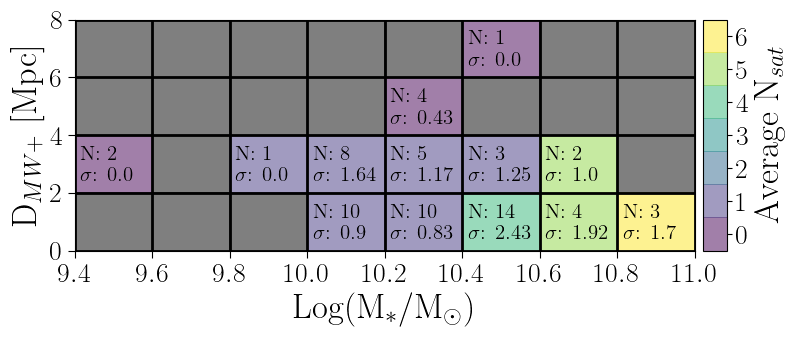}{.49\textwidth}{(a) $M_{\text{vir}}$ with simulated $R_{\text{vir}}$} 
    \fig{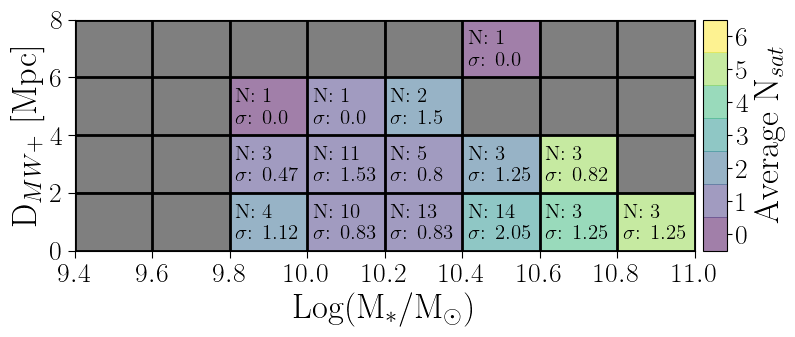}{.49\textwidth}{(b) SAGA II Comparison Sample}
}
\caption{The average number of satellites hosted by Milky Way analogs as a function of stellar mass and environment (distance to a Milky Way-sized or larger halo). The text in each box indicates the number of Milky Way analogs in that parameter space, as well as the standard deviation amongst the number of satellites. The left plot shows the $M_{\text{vir}}$ with simulated $R_{\text{vir}}$ sample, while the right plot shows the sample most analogous to the SAGA Survey. In both cases, the number of satellites appears to increase as stellar mass increases.}
\label{fig:boxplot}
\end{figure*}

\begin{figure*}
\gridline{
 \fig{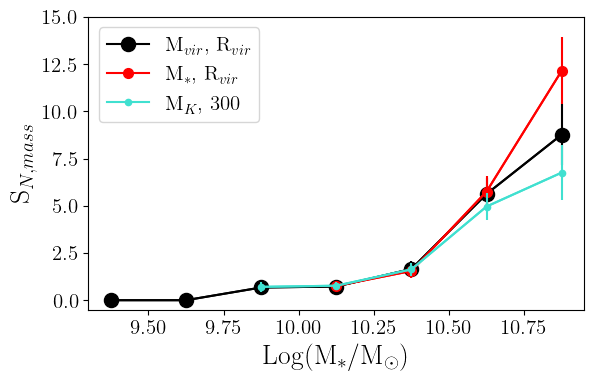}{.49\textwidth}{(a) Specific Frequency of $N_{\text{sat}}$ normalized to mass}
 \fig{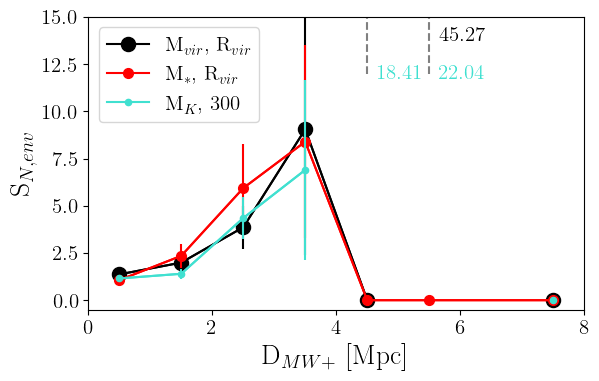}{.495\textwidth}{(b) Specific Frequency of $N_{\text{sat}}$ normalized to environment}
}
\caption{The specific frequencies of the number of satellites hosted by Milky Way analogs normalized to their (a) stellar mass and (b) distance to a Milky Way-sized or larger halo, $D_{MW+}$. The plots show the results for the $M_{\text{vir}}$ (black) and $M_*$ (red) with simulated $R_{\text{vir}}$ and SAGA II (blue) samples. Error bars represent the standard error within each bin ($\sigma/\sqrt{N}$). With the exception of some large outliers, the $S_{N,\text{env}}$ values do not show statistically significant trends. However, the $S_{N,\text{mass}}$ values show a clear positive trend.} 
\label{fig:SF}
\end{figure*}

\begin{figure*}
\gridline{
 \fig{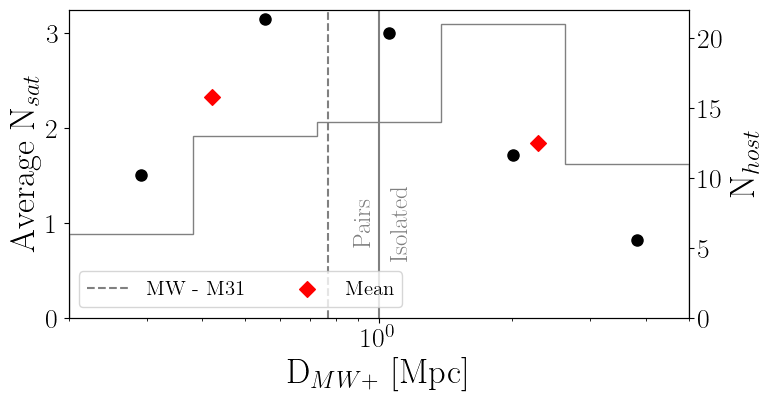}{.49\textwidth}{(a) $M_{\text{vir}}$ with simulated $R_{\text{vir}}$}
 \fig{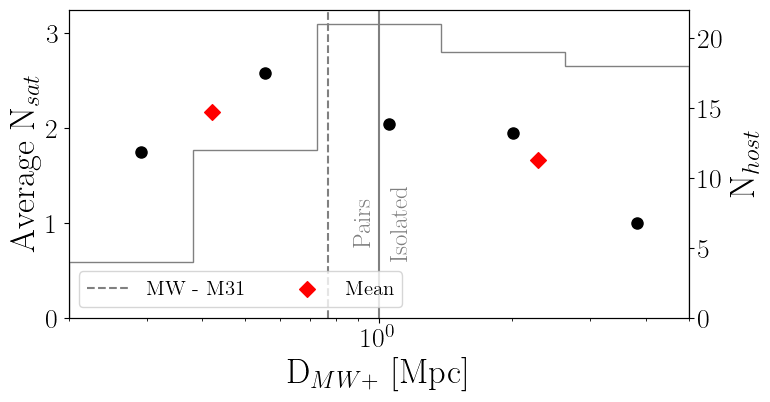}{.49\textwidth}{(b) SAGA II Comparison Sample}
}
\caption{Points show the average number of satellites hosted by Milky Way analogs binned by $D_{MW+}$ for our (a) $M_{\text{vir}}$ with simulated $R_{\text{vir}}$ and (b) SAGA II comparison sample. Hosts whose $D_{MW+}$ measurements are within 1 Mpc are considered to be in ``pairs'', while the rest are considered ``isolated'', and the means of these two subsamples are plotted as red diamonds. The histogram data shows the number of hosts in each bin, and the Milky Way-M31 distance is plotted as a dashed vertical line for reference. In both samples the isolated subset hosts fewer average satellites than the pairs, and this is driven predominantly by the most isolated bin.} 
\label{fig:NsatVsEnv}
\end{figure*}

\section{Results}
\label{sec:results}

Figure \ref{fig:lumfunc} shows the $V$-band satellite luminosity function for our sample of Milky Way analogs alongside data from the Milky Way and several Milky Way-like systems. The outer gray region outlines the space occupied by our $M_*$ within  $R_{\text{vir}}$ sample, while the black line and inner dark gray region indicate the mean and standard deviation. The Milky Way and M31 data are taken from \citet{SAGAI}. The NGC 4258 and NGC 4631 data were taken from \citet{Carlsten21b}, the M94 data from \citet{Smercina18}, and the M101 data from \citet{Bennet19}. We find that our sample of Milky Way analogs is in good agreement with these observations. We note that the space occupied by our sample remains largely unaffected when changing the Milky Way analog criteria.

\subsection{Host Effects on Satellite Accumulation}

In order to study how the physical properties of our Milky Way analogs affect their satellite populations, we separated our sample according to their mass and environment. Figure \ref{fig:boxplot} shows the average number of satellites hosted by the Milky Way analogs where the analogs are binned according to their stellar mass and minimum distance to a Milky Way-sized or larger halo, hereafter $D_{MW+}$. In calculating $D_{MW+}$, we consider the closest galaxy outside the system of the analog (i.e., not a satellite) that exceeds the minimum criteria of Milky Way analog under the given criteria. The text in each bin details \textit{N} (the number of analogs in that bin) and $\sigma$ (the standard deviation of the number of satellites hosted by analogs in that bin). A plot is shown for both our $M_{\text{vir}}$ with simulated $R_{\text{vir}}$ (left) and SAGA II comparison (right) samples. In all of our samples, the number of hosted satellites appears to increase with host mass, and potentially with decreasing $D_{MW+}$. However, this latter trend cannot be verified by eye as the box size of {\sc Romulus25} yields a lack of data in the upper regions of this plot (i.e., highly isolated hosts), so the apparent trend is not statistically significant.

While these macroscopic trends are present across all of our simulated samples, there are some notable differences in the distributions. We see that while the $M_{\text{vir}}$ definition includes analogs at a lower stellar mass, the number of analogs below $M_*=10^{10}$ $M_\odot$ is much larger in the SAGA II sample. Additionally, in the higher-mass bins there is fluctuation in both the number of analogs and hosted satellites due to the changing of the satellite selection radius from $R_{\text{vir}}$ to 300 kpc.

In an effort to quantify the ``by-eye'' trends seen in Figure \ref{fig:boxplot}, we looked at the specific frequency of the number of satellites hosted by our Milky Way analogs, $S_N$, normalized to their mass and environment. We use the following specific frequency equations adapted from \citet{Harris81}:
\begin{equation}
    S_{N,\text{env}} = N_{\text{sat}}\times10^{0.4(D-1.5)}
\end{equation}
\begin{equation}
    S_{N,\text{mass}} = N_{\text{sat}}\times10^{0.4(M-10.3)}
\end{equation}
Here, $N_{\text{sat}}$ is the number of satellites hosted by the Milky Way analog, $D$ is $D_{MW+}$ in units of Mpc, and $M$ is log($M_*$/$M_\odot$). The normalization values of 1.5 Mpc and 10.3 were chosen to be roughly the averages of the $M_*$ with simulated $R_{\text{vir}}$ sample. 

Figure \ref{fig:SF} shows the specific frequencies normalized to mass and environment for our $M_{\text{vir}}$ and $M_*$ with simulated $R_{\text{vir}}$ sample, as well as our SAGA II comparison sample. In looking at the trend with mass, the $S_N$ values consistently increase with the stellar masses of the Milky Way analogs. These results, which are present in all of our Milky Way analog samples, indicate that stellar mass exerts a large influence on satellite accumulation. The SAGA and ELVES surveys both observe this trend of satellite abundance increasing with host mass, though the trends they find are slightly weaker than ours (see Section \ref{sec:disc_rad} for discussion). Further, a study of seven nearby Milky Way-like systems with the Hyper Suprime-Cam on the Subaru telescope observes this trend as well \citep{Nashimoto22}. The trend of satellite abundance with host mass was also found by \citet{Font21} using the ARTEMIS suite of zoom-in simulations \citep{Font20}, and by \citet{Engler21} using the TNG50 simulation.

In looking at the trend with environment, Figure \ref{fig:SF}(b), we see some interesting behavior. The $S_N$ values increase somewhat linearly until $D_{MW+}\approx3.5$ Mpc, where future points go either to zero or extreme outliers. This would suggest that $N_{\text{sat}}$ increases as hosts become more isolated, but we note that a majority of our hosts ($\sim$60\%-70\%) have $D_{MW+}<$2 Mpc, so beyond this distance our samples get increasingly small, resulting in the large error bars and stochasticity of the higher-$D_{MW+}$ points. Thus, we see no definitive trend of satellite accumulation with environment, though one might become present with a larger sample of more isolated hosts. However, we do not believe that we can fully rule out an environmental impact on satellite accumulation through a measurement of specific frequency. Figure \ref{fig:NsatVsEnv} shows the average number of satellites hosted by our analogs when binned by their $D_{MW+}$ measurement. We see that if we split our analogs into subsamples of `pairs' and `isolated' based on having $D_{MW+}<1$ Mpc or $>1$ Mpc, the average numbers of hosted satellites, plotted as red diamonds, are notably different between the subsamples. This difference, which is present in all of our samples except $M_K$+Env. with simulated $R_{\text{vir}}$, is driven primarily by the low number of satellites hosted by analogs in the $\sim3-5$ Mpc bin. Although this is where our sample tapers off in $D_{MW+}$ space, we can see that the number of analogs in this bin is certainly nonnegligible, and may be hinting at a strong environmental impact on satellite accumulation in more extreme isolation. The stochasticity of our trends in Figure \ref{fig:SF}(b) compared to the more direct information from Figure \ref{fig:NsatVsEnv} leads us to believe that normalizing specific frequency to such an extremely variable parameter (in this case, large-scale environment) does not yield a reliable measurement.

\begin{figure}
    \gridline{
    \fig{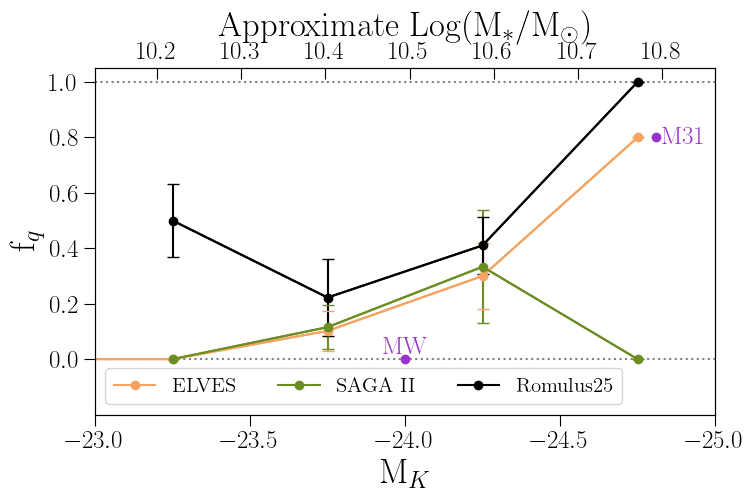}{.95\columnwidth}{(a) SAGA II Comparison Sample}}
    \gridline{
    \fig{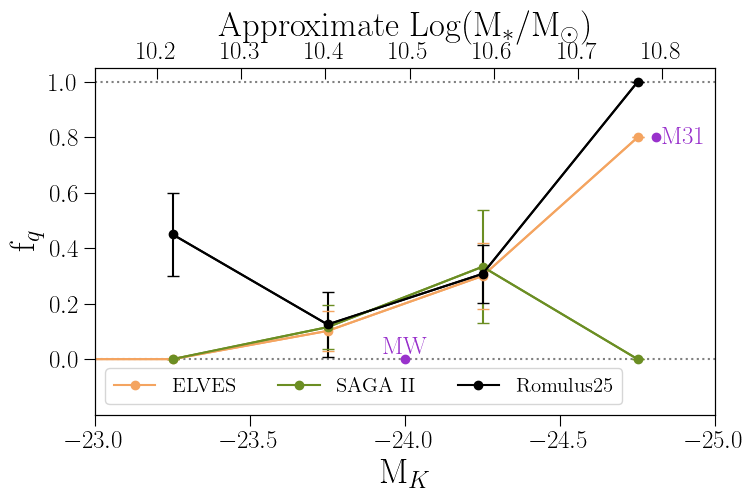}{.95\columnwidth}{(b) SAGA II Comparison Sample with Surface Brightness Cut}}
    \caption{(a) Quenched fraction plotted against $K$-band magnitude for the $M_K$+Env. sample from R{\sc omulus}25 compared to SAGA II \citep{SAGAII} and ELVES \citep{Carlsten22} data. (b) The same sample from R{\sc omulus}25 with addition criteria of requiring satellites to have $\mu_{\text{eff,}r}<25$ mag arcsecondond$^{-2}$. As a direct comparison, the SAGA and ELVES data plotted here only contain satellites with stellar masses above 10$^8$ $M_\odot$. Error bars represent the standard error within each bin. Approximate stellar mass values are taken from a linear fit between $M_K$ and log($M_*$/$M_\odot$) for our Milky Way analogs. The Milky Way and M31 values are taken from ELVES, and also only consider satellites with stellar masses above 10$^8$ $M_\odot$. While all three samples show the quenched fractions increasing with host brightness, our R{\sc omulus}25 sample exhibits slightly larger quenched fractions (particularly in the faintest bin) unless low surface brightness galaxies are removed. The SAGA and ELVES data are also in good agreement up to the brightest magnitude bin where the sample sizes are small.}
    \label{fig:SAGAComp}
\end{figure}

\begin{figure*}
\gridline{
 \fig{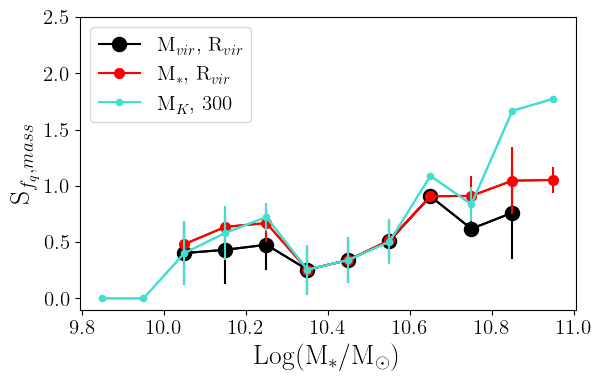}{.49\textwidth}{(a) Specific Frequency of $f_q$ normalized to mass}
 \fig{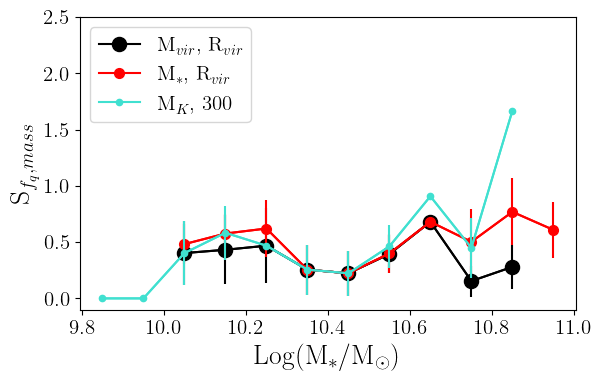}{.49\textwidth}{(b) Specific Frequency of $f_q$ normalized to mass with Surface Brightness Cut}
}
\caption{The specific frequencies of the quenched fraction of satellites hosted by Milky Way analogs normalized to their stellar mass. The plots show the results for the $M_{\text{vir}}$ (black) and $M_*$ (red) with simulated $R_{\text{vir}}$ and SAGA II (blue) samples. Subplot (b) applies the surface brightness cut to satellites ($\mu_{\text{eff,}r}<25$ mag arcsecondond$^{-2}$). Error bars represent the standard error within each bin. As with $N_{\text{sat}}$, there is a notable positive trend with Milky Way analog mass. However, applying the surface brightness cut to our satellites effectively removes the trend with mass.} 
\label{fig:SFq}
\end{figure*}

\begin{figure*}
\gridline{
 \fig{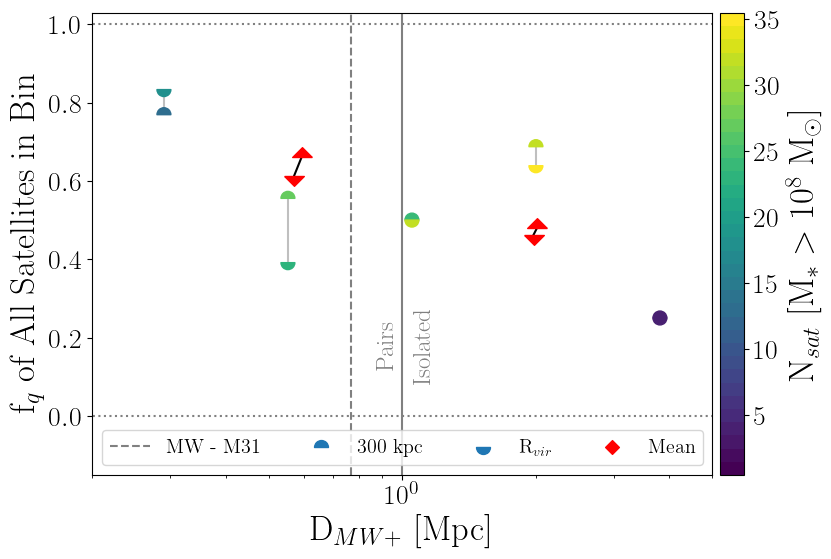}{.49\textwidth}{(a) $M_*$ Criteria}
 \fig{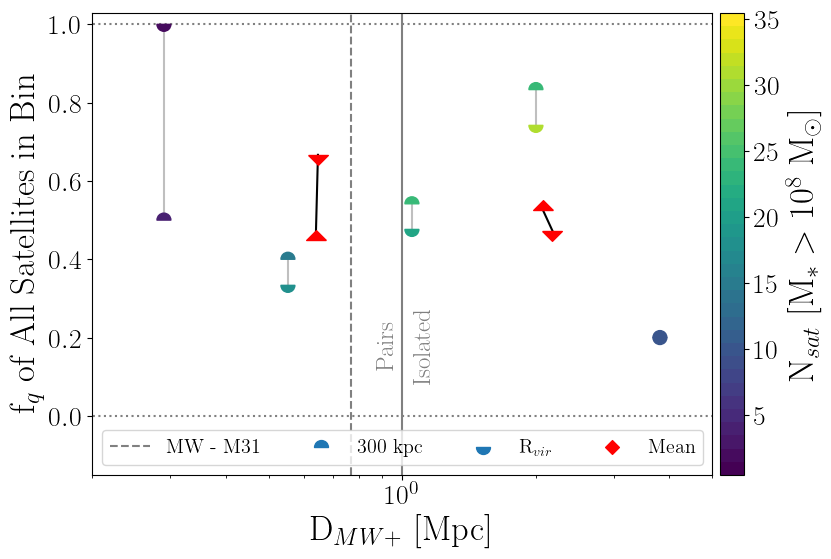}{.49\textwidth}{(b) $M_K$+Env. Criteria}
}
\gridline{
 \fig{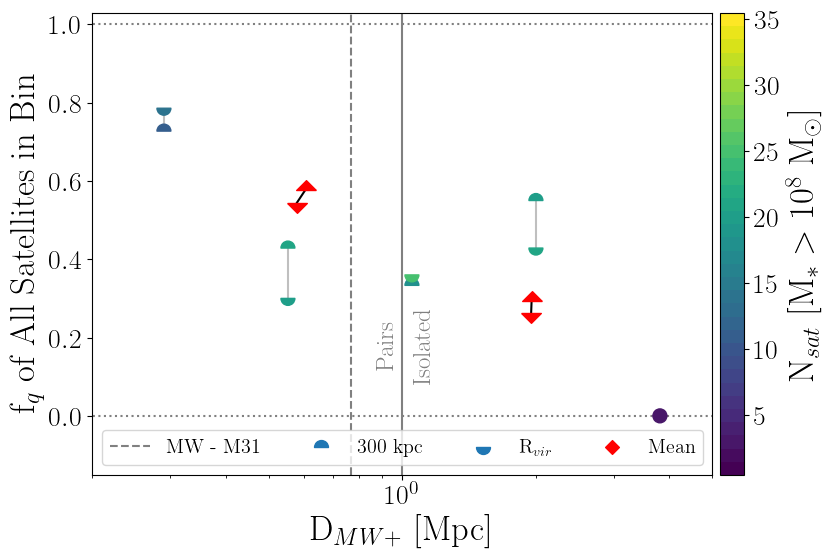}{.49\textwidth}{(c) $M_*$ Criteria with SB cut}
 \fig{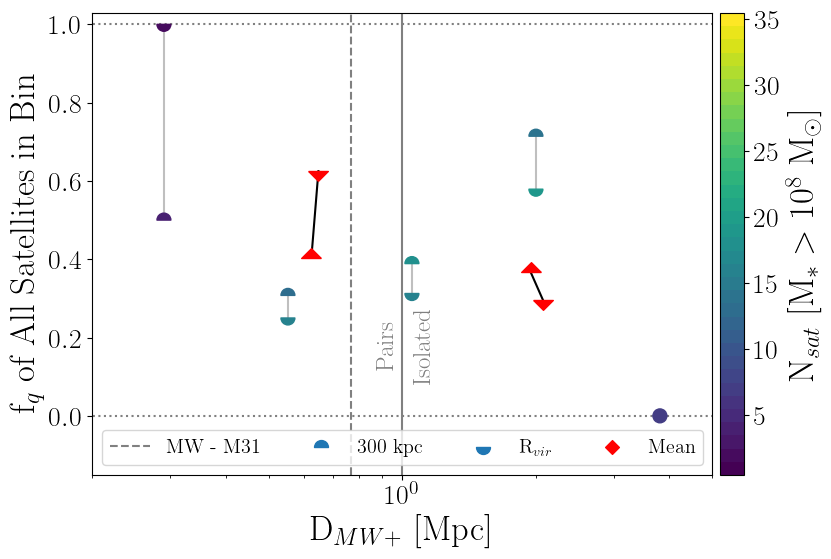}{.49\textwidth}{(d) $M_K$+Env. Criteria with SB cut}
}
\caption{Quenched fraction plotted against the distance to the closest Milky Way halo or larger for the (a) $M_*$ and (b) $M_K$+Env. criteria. Analogs are binned by $D_{MW+}$ and the quenched fractions are taken from an aggregate of all satellites in each bin. The upper-hemisphere points represent analogs from the sample with a Milky Way radius of 300 kpc, while lower-hemisphere points represent the sample using the simulated virial radius with vertical lines bridging the two. The colors of each point represent the number of hosted satellites with stellar mass greater than 10$^8$ $M_\odot$. The samples are separated into ``Pairs'' and ``Isolated'' by whether the closest Milky Way or larger halo is within 1 Mpc, and the means of each sample are denoted by the red diamonds. For reference, the Milky Way-M31 distance is plotted with a vertical dashed line. Subplots (c) and (d) require satellites to have $\mu_{\text{eff,}r}<25$ mag arcsecondond$^{-2}$. The pairs exhibit a slightly higher mean quenched fraction, and changing the satellite selection radius from 300 kpc to $R_{\text{vir}}$ typically takes the quenched fraction to an equivalent or lower value. Applying the surface brightness cut to our satellites strengthens the disparity between paired and isolated samples.}
\label{fig:QvE}
\end{figure*}

\subsection{Host Effects on Satellite Quenching}
\label{sec:res_q}

In addition to studying the number of satellites hosted by our analogs, we also analyzed the quenched fraction of the satellites. When studying the quenched fraction ($f{\rm_q}$), we only consider satellites with a stellar mass of at least $10^8$ $M_\odot$, as R{\sc omulus}25 may be subject to numerical overquenching below this mass \citep{Wright21}. A galaxy is considered quenched if its instantaneous specific star formation rate (sSFR) is below $10^{-11}\text{ yr}^{-1}$. The instantaneous sSFR is calculated using the expected SFR from gas particles meeting the temperature and density thresholds for SF given in Section \ref{sec:sf}. In Figure \ref{fig:SAGAComp}, we show our quenched fractions as a function of host $K$-band magnitude for the $M_K$+Env.~with 300 kpc satellite selection (our SAGA II comparison sample), and compare our results to data from the SAGA and ELVES surveys \citep{SAGAII,Carlsten22}. For a direct comparison, we only consider SAGA and ELVES satellites with stellar masses above 10$^8$ $M_\odot$. We note, however, that the SAGA and ELVES surveys' methods of determining quenching are different than ours: SAGA considers a satellite quenched if it lacks strong H$\alpha$ emission (equivalent width (EW) of H$\alpha$$<2$\AA) and ELVES considers a satellite quenched if it exhibits an early-type morphology, i.e., not exhibiting clear star-forming structures such as blue clumps or dust lanes (this correlates with color as well; see \citet{Carlsten21a} for an in-depth discussion).  Our sSFR quenched definition was shown \citep[see][]{Sharma23} to yield a good match to galaxies identified observationally as quenched using EW[H$\alpha$]$<2$\AA ~and $D_{\rm n}4000 > 0.6 + 0.1\log_{\rm 10}M_*$ \citep[as in][]{Geha2012}.

While all three samples show quenched fractions increasing with host brightness, our simulated sample exhibits slightly larger quenched fractions than the observational surveys, with the exception of the lowest-mass bin where the difference becomes significant (see Section \ref{sec:disc_qf} for discussion). The SAGA and ELVES data are in very good agreement up to the brightest magnitude bin, where the sample sizes are only one host for ELVES (M31) and two hosts for SAGA (NGC 5792 and NGC 7541). This agreement within the high-mass satellite subset is interesting, as the SAGA and ELVES quenched fractions are quite different when considering their full samples. \citet{Carlsten22} find that the quenched fractions of the Local Volume are significantly higher than the SAGA sample (their figures 11 \& 12), particularly in the low-mass satellite regime. In Figure \ref{fig:HostComp}(a), we see that the ELVES survey contains a much larger number of faint satellites when compared to SAGA, but also that ELVES hosts (along with those of R{\sc omulus}25) probe fainter magnitudes. 

In studying the ARTEMIS simulations, \citet{Font22} found that the SAGA detection methods may be preferentially selecting star forming or recently quenched satellites near their completeness limit, missing a notable population of quenched dwarfs. This detection bias could explain the difference between SAGA and ELVES low-mass satellites, both the abundances and quenched fractions. Following \citet{Font22}, in Figure \ref{fig:HostComp}(b) we apply an additional cut to our SAGA II comparison sample by requiring satellites to have $\mu_{\text{eff,}r}<25$ mag arcsecond$^{-2}$. As in the ARTEMIS simulations, we find that this cut lowers the resultant quenched fractions, and brings our results (particularly the middle bins) into excellent agreement with SAGA and ELVES.

To quantify the trend of quenched fraction with mass seen in Figure \ref{fig:SAGAComp}, and to search for a trend with environment, we again used the specific frequency in Equation (2) with $N_{\text{sat}}$ replaced by $f_{\text{q}}$. Figure \ref{fig:SFq}(a) shows our quenched fraction specific frequencies for the $M_{\text{vir}}$ and $M_*$ with simulated $R_{\text{vir}}$ samples, as well as our SAGA II comparison sample. We find that, as with the number of hosted satellites, we see a trend of $S_N$ with host mass, indicating that larger hosts are expected to yield higher quenched fractions. However, we note that this trend is not as strong as the one seen in Figure \ref{fig:SF}(a). Interestingly, when applying the satellite surface brightness criteria in Figure \ref{fig:SFq}(b), we see that our trend of quenched fraction with host mass is effectively erased. As the high-mass end of Figure \ref{fig:SFq}(b) is strongly affected by this surface brightness cut, it seems that the preferentially quenched satellites below this threshold are more common in higher-mass hosts, which is consistent with Figure \ref{fig:SFq}(a) implying a larger number of quenched galaxies in this regime. Our results agree with \citet{Engler23} who, using the TNG50 run from the IllustrisTNG simulations, found that massive hosts exhibit systematically larger satellite quenched fractions. Further, \citet{Engler23} found no difference in quenching between isolated and paired analogs when considering satellites within 300 kpc of their host (see Section \ref{sec:disc_pair} for discussion).

To look for a trend with environment, we examined the quenched fractions of our systems plotted against $D_{MW+}$. Figure \ref{fig:QvE} bins our analogs in $D_{MW+}$ space, and shows the quenched fraction of all satellites hosted by analogs in each bin. Again, pairs are identified as having $D_{MW+}< 1$ Mpc. The figure shows data for our $M_*$ and  $M_K$+Env.~samples (both $R_{\text{vir}}$ and 300 kpc), where the color of each point represents the number of satellites in each bin. We find that the average quenched fraction is higher among pairs than isolated analogs, though the magnitude of this difference is not ubiquitous across our samples (see Section \ref{sec:disc_pair} for discussion). In fact, while not shown in the plot, the $M_{\text{vir}}$ criteria exhibit no notable difference in the average quenched fractions of paired and isolated analogs. We also find that in the switch from 300 kpc to $R_{\text{vir}}$ when identifying satellites, hosts typically have either the same or lower quenched fractions and a higher satellite count. This indicates that restricting satellites to within 300 kpc for this host range is more likely to exclude satellites, and that the satellites beyond 300 kpc are predominantly star forming; though this is still only when considering satellites with $M_*>10^8$ $M_\odot$. In applying the surface brightness cut in subplots (c) and (d) we find that, while the resultant averaged quenched fractions are lower, the difference between the isolated group and pairs notably increases in all samples.

Similar to Figure \ref{fig:NsatVsEnv}, this difference between our `pairs' and `isolated' subsamples is largely driven by the most isolated bin, where the quenched fractions are extremely low. This, again, could be alluding to a strong environmental effect on satellite quenching in the highly isolated regime that we do not quite capture in R{\sc omulus}25. This result is at odds with the aforementioned study of TNG50 by \citet{Engler23}, as well as the study of FIRE-2 by \citet{Samuel22}, both of whom find no difference in quenching between isolated and paired analogs. However, the simulations studied in \citet{Samuel22} are zoom-in simulations, and thus would not capture the large-scale isolation in which we begin to see differences between paired and isolated hosts.

While this work does not consider signs of conformity within satellites, we did perform some cursory analysis on how the quenched fractions and number of hosted satellites relate to the instantaneous SFRs of our analogs. We find that the quenched fractions exhibit no notable trend with host SFR. In looking at the average number of hosted satellites, though, we do see a correlation of more populated systems having a higher host SFR. However, we believe this is just a reflection of the number satellites increasing with host mass, as scaling to sSFR largely removes this correlation.

\section{Discussion}
\label{sec:discussion}
In Section \ref{sec:sample}, we discussed the various methods by which we identified Milky Way analogs and satellites. While shifting between these definitions has no effect on our conclusions, there are subtle impacts worth noting.

\subsection{Satellites within $R_{\text{vir}}$ versus 300 kpc}
\label{sec:disc_rad}
In Figures \ref{fig:boxplot} \& \ref{fig:SF} we showed host stellar mass to be a driving factor in satellite accumulation, but this trend is less prominent when using our SAGA II comparison sample. This appears to be the result of using 300 kpc to identify satellites, not the selection on $K$-band magnitude, as our $M_K$+Env.~with $R_{\text{vir}}$ sample actually exhibits the strongest trend. In fact, identifying satellites via a 300 kpc selection rather than $R_{\text{vir}}$ reduces the strength of the mass trend in \text{all} criteria (though the trend is still prominent). The weakening of the trends is the result of analogs in the high-mass regime (where the trends manifest), which have virial radii larger than 300 kpc and exclude satellites in this shift to 300 kpc. This shift is in agreement with the ARTEMIS simulations \citep{Font21}, in which satellite abundance trends strongly with host mass, but the trend is weakened when SAGA observation selection criteria are applied  ($M_{r,\text{sat}}<-12,\  \mu_{\text{eff},r}<25$ mag arcsecond$^{-2}$, and within 300 kpc of the host).

When considering quenched fractions, our choice of satellite selection radius also seems to have a noticeable effect on our $M_K$+Env. sample. In Figure \ref{fig:QvE}(b), the switch from $R_{\text{vir}}$ to 300 kpc typically raises the quenched fraction while lowering the number of satellites (with the notable exception of the least isolated bin). Thus, within the context of satellites with $M_*>10^8$ $M_\odot$, it seems applying a satellite cut of 300 kpc to the $K$-band magnitude analog selection primarily removes star-forming satellites from massive hosts, and biasing the global quenched fraction high. 

Since the 300 kpc selection results in a more centrally located satellite population, it is likely that these satellites had an earlier infall time and underwent more ram pressure stripping when compared to satellites near or beyond 300 kpc from the host. This effect is present in Figure \ref{fig:SFq} as well, wherein the SAGA II comparison sample exhibits the strongest trend of quenched fraction with host mass.  These results are consistent with those from the TNG50 simulation \citep{Engler23}, another large-volume, uniform-resolution simulation with comparable resolution to R{\sc omulus}25.

\subsection{Quenched Fraction Discrepancy}
\label{sec:disc_qf}
The shift from $R_{\text{vir}}$ to 300 kpc, however, does not explain why our quenched fractions are higher than those of SAGA and ELVES (Figure \ref{fig:SAGAComp}). Our SAGA II comparison sample uses 300 kpc as a selection radius, and our results indicate that if SAGA and ELVES had access to the virial radii of their hosts, their quenched fractions would be lower.  \citet{Donnari21} find that the adopted definition of quenching and using 2D projected distances can both affect the resultant quenched fractions. Notably, the quenched fractions of R{\sc omulus}25 are in better agreement with the observations when satellites with $\mu_{\text{eff,}r}<25$ mag arcsecond$^{-2}$ are removed, in agreement with \citet{Font22}.  The exception is the faintest bin, where a key factor may be the resolution of R{\sc omulus}25.  The lower resolution of the volume is unable to resolve a multiphase ISM, i.e., there is no extremely dense gas \citep[][and references within]{Tremmel19,Tremmel20}. Thus, all of the gas is ``puffy'' and overly susceptible to ram pressure stripping and quenching.

\citet{Dickey21} found that large-scale cosmological simulations overquench isolated galaxies below $M_*=10^9$ $M_\odot$ when compared to the Sloan Digital Sky Survey. The authors attribute this to overefficient feedback, which is typically tuned to recreate quenched fractions found in the Local Volume. In looking at R{\sc omulus}25, \citet{Sharma23}, also found that isolated dwarfs exhibit a higher quiescent fraction when compared to observations, but that this can be entirely attributed to the presence of massive black holes and their feedback. Although we are not studying isolated dwarfs in this work, it is still likely that these feedback properties are influencing our results. We note, however, that there are only six satellites in our SAGA II comparison sample with black holes and $M_*>10^8$ $M_\odot$ so this does not notably affect our results. 

\subsection{Isolated vs. Paired Hosts}
\label{sec:disc_pair}
Figure \ref{fig:QvE} suggests that pairs exhibit higher quenched fractions than more isolated analogs, but there are some caveats preventing us from making a more robust statement about the environment's effect on the quenched fraction. First, we are only considering satellites with $M_*>10^8$ $M_\odot$. Within our SAGA II sample, this is only $\sim$56\% of our total satellite population and they are hosted by $\sim$72\% of our Milky Way analogs with a nonzero satellite count (or $\sim$49\% of all Milky Way analogs), so a large section of our population is being removed. Secondly, our simulation box size prevents us from having a large sample of highly isolated analogs; only $\sim14\%$ of our SAGA II sample analogs have $D_{MW}>3$ Mpc. Finally, by ignoring low-mass satellites, we are looking at the quenched fractions of several systems with few satellites (only one or two satellites). Around 43\% of the high-mass satellite-hosting analogs in our SAGA II sample contain only one satellite above our resolution limit, so their quenched fractions can only occupy the extremes of 0 and 1, and in Figures \ref{fig:SFq} \& \ref{fig:QvE} these systems are being considered equally alongside systems with as many as eight high-mass satellites (though the binning in Figure \ref{fig:QvE} should alleviate this issue). These combined effects yield a sample that is lacking low-mass satellites (and thus the analogs' full satellite distributions) as well as highly isolated hosts, making it difficult for us to extrapolate our results to the Universe at large.

Recently, \citet[][TNG50]{Engler23} and \citet[][FIRE-2]{Samuel22} found no difference in the satellite quenched fractions of paired and isolated hosts in their simulations. Further, \citet[][FIRE-2]{GarrisonKimmel19b} found that the satellites of isolated Milky Way-mass galaxies have nearly identical SF histories to satellites of Milky Way analogs in Local Group-like pairs. However, these results were only considering satellites within 300 kpc of the host. In looking further out to 300-1000 kpc, \citet{Engler23} find that paired, Local Group-like hosts exhibit significantly larger quenched fractions than their isolated counterparts.

\section{Conclusions}
\label{sec:conclusions}
Using the R{\sc omulus}25 simulation, we have created various samples of Milky Way analogs along with their satellite distributions. We explored the role of host mass and environment on satellite numbers and quenched fractions. Our results can be summarized as follows.
\begin{itemize}
    \item When testing various criteria for defining a Milky Way analog, from more theoretically motivated ($M_{\text{vir}}$) to more observationally motivated ($M_*$ and SAGA-like), we find that the resultant samples do not fully overlap. Within the overlapping regions, galaxies may also be defined as analogs in one sample but not in another due to environmental criteria (see Table \ref{tab:sample} and Figure \ref{fig:defcomp}).
    \item The number of satellites hosted by a Milky Way analog increases predominantly with host stellar mass, while environment may have a significant impact in high isolation ($D_{MW+}>3$ Mpc; see Figures \ref{fig:boxplot}-\ref{fig:NsatVsEnv}).
    \item The quenched fraction (for satellites with $M_*>10^8$ $M_\odot$) of our analogs increases with host mass (see Figures \ref{fig:SAGAComp}(a) \& \ref{fig:SFq}(b)), but applying a surface brightness cut to satellites can erase this trend (see Figure \ref{fig:SFq}(d)).
    \item Being in a pair may yield higher satellite quenched fractions, but it is hard to draw statistically robust results given the small volume of {\sc Romulus25} and the fact that we can only study satellites down to $M_*=10^8$ $M_\odot$ to avoid numerical overquenching. (see Figure \ref{fig:QvE}).
\end{itemize}

We find that the distributions of both the Milky Way and M31 are well explained by our sample, with M31 being at the highly populated edge of our sample space. This is in agreement with the SAGA and ELVES surveys, where ELVES found the Local Volume to be slightly more populated and exhibiting a steeper luminosity function when compared to the full SAGA sample. Additionally, we are in agreement with ELVES in finding that the stellar mass of a Milky Way analog seems to be the dominant factor in both the number of hosted satellites and the number of quenched satellites. Interestingly, in our study of quenching, we find that the SAGA and ELVES results are in good agreement for satellites with $M_*>10^8$ $M_\odot$, suggesting that their discrepancy in quenched fraction comes from lower-mass satellites, which we are unable to probe here due to numerical effects that artificially quench simulated galaxies.  
However, our results support the notion put forward in \citet{Font22} that SAGA is missing a large population of low surface brightness satellites near its detection limit that are preferentially quenched.

\begin{acknowledgments}
J.D.V. is supported by the Homer L. Dodge Fellowship from the University of Oklahoma. Long before the University of Oklahoma was established, the land on which the University now resides was the traditional home of the “Hasinais” Caddo Nation and Kirikiris Wichita \& Affiliated Tribes; more information can be found \href{https://www.ou.edu/cas/nas/land-acknowledgement-statement}{here}.  C.C. was supported by the NSF under CAREER grant AST-1848107. A.M.B. was partially supported by NSF grant AST-1813871. M.T. was supported by an NSF Astronomy and Astrophysics Postdoctoral Fellowship under award AST-2001810. The Romulus simulations are part of the Blue Waters Sustained Petascale Computing project, which is supported by the National Science Foundation (awards OCI-0725070 and ACI-1238993) and the state of Illinois. Blue Waters is a joint effort of the University of Illinois at Urbana–Champaign and its National Center for Supercomputing Applications. This work is also part 12 of a Petascale Computing Resource Allocations allocation support by the National Science Foundation (award number OAC-1613674). This work also used the Extreme Science and Engineering Discovery Environment (XSEDE), which is supported by National Science Foundation grant No. ACI-1548562.
\end{acknowledgments}

\section*{Data Availability}
The data for this work were generated from a proprietary branch of the {\sc ChaNGa} N-Body+SPH code \citep{Changa}. The public repository for {\sc ChaNGa} is available on github https://github.com/N-BodyShop/changa). Analysis was conducted using the publicly available softwares pynbody \citep[][https://github.com/pynbody/pynbody]{pynbody} and TANGOS \citep[][https://github.com/pynbody/tangos]{Tangos}. These results were generated from the \textsc{Romulus25} cosmological simulation. The raw output from this simulation can be accessed upon request from Michael Tremmel (mtremmel@ucc.ie), along with the TANGOS database files that were generated from these outputs and directly used for this analysis.

\bibliography{bibliography.bib}
\end{document}